\documentclass[apj]{emulateapj}
\usepackage{graphics}
\usepackage{natbib}
\usepackage{color}

\newcommand{\as}{$^{\prime\prime}$ }

\newcommand{\nh}{$N_{\rm{H}}$}
\newcommand{\nustar}{\textit{NuSTAR}}
\newcommand{\xmm}{\textit{XMM-Newton}}
\newcommand{\chandra}{\textit{Chandra}}
\newcommand{\suzaku}{\textit{Suzaku}}

\newcommand{\swift}{\textit{Swift}}
\newcommand{\rxte}{\textit{RXTE}}
\newcommand{\degree}{$^\circ$}

\begin{document}

\title{Measurement of the absolute Crab flux with \textit{NuSTAR}.}
\author{Kristin K. Madsen$^1$, Karl Forster$^1$, Brian W. Grefenstette$^1$, Fiona A. Harrison$^1$, and Daniel Stern$^2$}
\affiliation{$^1$ Cahill Center for Astronomy and Astrophysics, California Institute of Technology, Pasadena, CA 91125, USA\\
$^2$ Jet Propulsion Laboratory, California Institute of Technology, Pasadena, CA 91109, USA\\
}

\begin{abstract}
We present results from a \nustar\ observation of the Crab made at a large off-axis angle of 1.5\degree. At these angles X-rays do not pass through the optics, but rather illuminate the detectors directly due to incomplete baffling. Due to the simplicity of the instrument response in this configuration and the good absolute calibration of the detectors, we are able to measure the absolute intrinsic flux of the Crab to better than 4\%. We find the spectral parameters of the powerlaw to be $\Gamma=2.106\pm 0.006$, $N=9.71\pm 0.16$, in agreement with the values measured 42\,years ago by \citet{Toor1974}. This suggests that the observed variability of the Crab is not part of a long term trend, but instead results from fluctuations around a steady mean. The \nustar\ observation also enabled improved measurement of the detector absorption parameters without the added complications of the mirror response.
\end{abstract}


\section{Introduction}
The Crab is the iconic plerionic pulsar wind nebula (PWN), characterized by a center-filled synchrotron nebula that is powered by a magnetized wind of charged particles emanating from a centrally located pulsar formed during the supernova explosion \citep{Weiler1978,KC19841}. Its phase-averaged spectral shape in X-rays (nebula + pulsar) can be approximated by a powerlaw, $dN/dE = N\,E^{-\Gamma}$ photon cm$^{-2}$\,s$^{-1}$\,keV$^{-1}$, and the absolute flux and stability of the Crab in the X-ray band has been an intense topic of research. Numerous balloon and rocket-borne instruments flying proportional, Geiger, and scintillation counters were built to address this topic; a full list can be found in \citet{Toor1974}. The combined powerlaw fit to all these data gave a normalization of $N=9.5$ and $\Gamma=2.08\pm 0.05$ (the error on the normalization was included in the index), to an estimated precision in flux of  $\pm 15\%$ at 2--10\,keV and $\sim50\%$ at 10--70\,keV. \citet{Toor1974} compared this to their own rocket experiment, which flew a set of 10 proportional counters, and obtained a spectrum over 2--60\,keV with a best fit of $\Gamma=2.10\pm 0.03$ and $N=9.7\pm 1.0$. They concluded that, to within 10\%, the Crab was a steady source and well-suited as a calibration target for X-ray instrumentation. 

Since that time the Crab has been extensively used for exactly that purpose. However, the actual values of $N$ and $\Gamma$, to which instruments should calibrate, have remained ambiguous, and it is debatable whether a powerlaw is truly representative of the phase averaged integrated spectrum in the X-ray band \citep{Weisskopf2010}; a spatial breakdown of the spectrum has shown the continuum to vary across the nebula \citep{Mori2004, Madsen2015b}, and the pulsed spectrum to be best represented by a logarithmic parabolic powerlaw \citep{Kuiper2001, Madsen2015b}, which combined should not add up to another powerlaw. Furthermore, over a three year period (2008--2010) the Crab decreased its overall flux from the beginning of the observation by $\sim 3.5\%$\,yr$^{-1}$ \citep{Wilson2011}. In the same period the long term lightcurve shows that the Crab goes through variations on a yearly timescale with accompanying slope changes of a few percent \citep{Shaposhnikov2012}. 

Measurements of the Crab done in the last few decades by space-borne observatories is summarized by \citet{Kirsch2005} and shows that while there is agreement in the parameter space the spread is still large. In the energy range of interest (3--50\,keV), slope values span $\Gamma=2.05-2.13$ with normalizations of $N=7-11$. These differences are likely due to a conglomeration of instrumental challenges in sensitive low-energy observatories to high flux rates, flux variations in the source itself, and calibration differences; many observatories, including \nustar, have calibrated their instrument response against a set of spectral parameters that instrument teams have individually assumed for the Crab. For example, \nustar\ was calibrated against $\Gamma=2.1$, $N=8.7$, and \nh=$2.2 \times 10^{21}\mathrm{cm}^{-2}$ \citep{Madsen2015}, while \textit{RXTE}/PCA against $\Gamma=2.11$, $N=11$, and \nh=$3.4 \times 10^{21}\mathrm{cm}^{-2}$ \citep{Shaposhnikov2012}. 

It is generally agreed that collimated instruments are easier to absolutely calibrate, but all such observatories from recent times have in some manner been calibrated against the Crab.
We present here a new measurement of the instantaneous absolute Crab flux, where we have made use of the very simple stray-light geometry to circumvent the optics on-board \nustar. By using ground calibrated detector responses only, which are known to 1\% above 5\,keV, we can measure the Crab flux to better than 4\%.

\section{NuSTAR as a collimator}
The \textit{Nuclear Spectroscopic Telescope Array} (\nustar) is a focusing X-ray observatory operating in the 3--79\,keV band. It carries two co-aligned focusing X-ray optics matched to two identical Focal Plane Modules (FPMA and FPMB), each composed of four solid state CdZnTe pixel detectors (enumerated Det0 through Det3). The optics and FPMs are separated by a 10.15\,m unshrouded mast. More detailed information on the observatory can be found in \citet{Harrison2013}.

\begin{figure}
\begin{center}
\includegraphics[width=0.5\textwidth]{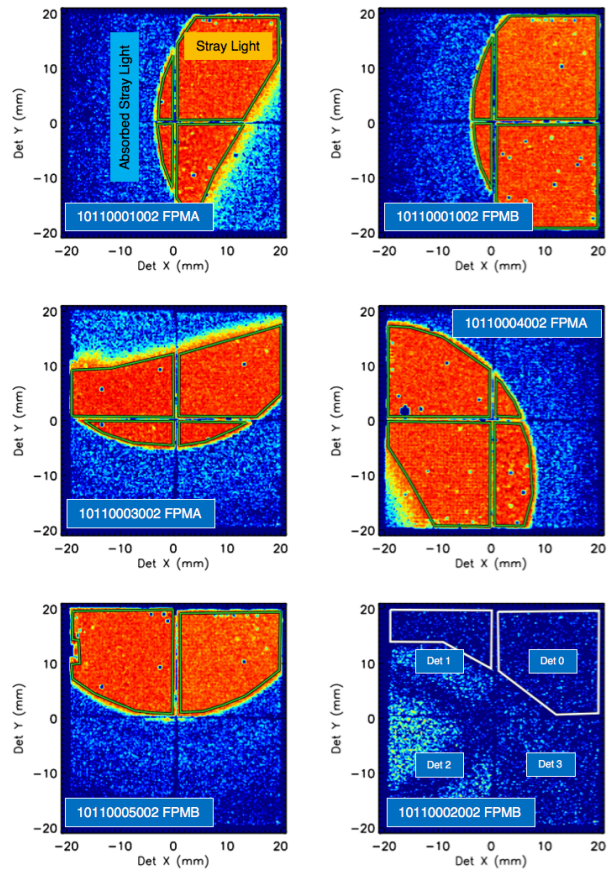}
\end{center}
\caption{Contour color plots of the detectors with logarithmic scaling. Green and white polygons show extraction regions. As shown in the bottom right, the individual detectors are enumerated counter clockwise from 0 to 3. The detector dimensions are given in mm.}
\label{detplots}
\end{figure}

``Stray-light" in \nustar\ is the term used to describe light that enters through the detector apertures without being reflected through the X-ray optics. The open geometry of the unshrouded mast allows light to enter unobstructed and reach the focal plane at angles of $\sim$ 1--5\degree, essentially turning \nustar\ into a collimated instrument. The triangular shape of the optical bench determines the smallest angle, while the radius of the aperture determines the largest allowable angle through which stray-light can enter. This causes the stray-light to appear as shown in Figure \ref{detplots} with a circular edge due to the aperture stop opening. The angular cut-away of some of the stray-light regions is the obscuration of the optical bench. 

Typically, stray-light from bright sources is not desirable. As part of standard operations the \nustar\ Science Operations Center avoids, whenever possible, observations that cause stray-light to appear at the location of a focused source. However, these Crab observations, listed in Table \ref{obsid}, were designed with the specific intent of getting as much stray-light as possible for the dual purpose of getting an independent measurement of the Crab spectrum and flux, and measuring the detector absorption parameters, which affect the instrument response below 5\,keV, of the 8 individual detectors without the added complication of the mirror response.

The detector absorption comes from a Pt contact coating on the surface, and a CdZnTe dead layer. The thickness of this layer was initially calibrated after launch in 2012 using 3C\,273 and the Crab \citep{Madsen2015}, but because of a degeneracy with the mirror effective area, the two effects could not be clearly separated in the analysis. By eliminating the optics response using the Crab observations reported here, the two can be separated.

\begin{table*}
\centering
\caption{Crab Stray-light Observations}
\begin{tabular}{l||c|c|c|c|c|c|c}
\hline
obsID & Date & Pointing RA & Pointing Dec & PA & Exposure & FPMA & FPMB\\
 & (Year:DoY)  & (deg) & (deg) & (deg) & (ks) & (det \#) & (det \#)\\
\hline
10110001002 & 2015:290 & 85.0310 & 22.8145 & 154 & 21.1 & 0,1,2,3 & 0,1,2,3 \\
10110002002 & 2015:291 & 81.9331 & 21.2145 & 154 & 20.1 & bkg\footnote{Used for background.} & nuskybg\footnote{Used with \texttt{nuskybkg} to obtain background.} \\
10110003002 & 2015:291 & 82.8331 & 23.4145 & 154 & 22.8 & bkg & 0,1,2,3 \\
10110004002 & 2016:92 & 84.6331 & 21.1154 & 333 & 20.9 & 0,1,2,3 & bkg \\
10110005001 & 2016:93 & 84.5331 & 20.3145 & 333 & 21.8 & bkg & 0,1 \\
10002001009\footnote{Focused on-axis observation.} & 2016:92 & 83.6331 & 22.0145 & 333 & 5.2 & - & - \\
\hline
\end{tabular}
\label{obsid}
\end{table*}

\section{Data Reduction}
There are in total five stray-light observations and one focused. For the stray-light observations, the Crab was placed $\sim 1.5$\degree\ off-axis at different RA and Dec locations (see Table \ref{obsid}), which combined with the observatory Position Angle (PA) determines the stray-light pattern. Because of the relatively large angles, compared to the pointing stability, under which the stray-light arrives the stray-light patterns are very reproducible and insensitive to small pointing errors. The first three observations were done in October 2015 and the last two, along with the focused observation, in April 2016. 

The incident countrate of the Crab is $\sim 2$\,photon\,s$^{-1}$\,cm$^{-2}$ in the energy range 3--80\,keV. On average, the stray-light covered two detectors of 4\,cm$^{2}$ each, resulting in a maximum of $\sim 16$ counts s$^{-1}$. The data were reduced using the \texttt{NuSTARDAS} v1.6.0 pipeline procedure \texttt{nupipeline} with calibration data base (CALDB) version 20160502, although with the updated gain file from  CALDB version 20160606. We used default parameter settings, but had to apply additional background filters for 10110003002 due to a bright solar flare. We used settings \texttt{SAAMODE=optimized} and \texttt{TENTACLE=yes}. We also had to remove by hand a background solar spike from 10110002002, which was not removed with any of the available background filtering settings. 

We did not use \texttt{nuproducts} to extract spectra, but designed custom code to operate directly on the cleaned event-list in detector coordinates. We extracted the spectra in these coordinates rather than sky coordinates because it is a natural frame for a ``collimated" telescope and makes the calculation of area trivial. Also, since the aperture stop is fixed with respect to the focal plane modules, the edges of the stray-light region are sharper. We extracted the spectra using the green polygons, as shown in Figure \ref{detplots}, and obtained one spectrum per detector (four detectors per module) per observation (Table \ref{obsid} lists the illuminated detectors for each observation). We combined all the spectra from the same detector in the same module for epoch 2015 and 2016 separately; e.g. we obtain one spectrum for Det0 FPMA and one for FPMB for epochs 2015 and 2016. In this manner, we end up with eight spectra for epoch 2015 (Det0A, Det1A, Det2A, Det3A, Det0B, Det1B, Det2B, and Det3B); but only six spectra for 2016 (Det0A, Det1A, Det2A, Det3A, Det0B, and Det1B) since Det2B and Det3B were not illuminated by stray-light during this epoch.

The instrument response for stray-light is simple and consists of the detector redistribution matrix (RMF), the illuminated area of the detector, and the absorption components in the path of the stray-light. The individual RMFs for each detector are directly available from the \nustar\ CALDB, and the detector area is easily calculated as the area of the polygon used for the extraction, minus the area covered by dead pixels. The dead pixel list is available from the bad pixel file. The only source of absorption, apart from the absorbing detector layer which we fit for, is the Be window and it has a thickness of 100$\mu$m with a throughput of 92\% at 5\,keV and 98\% at 10\,keV. We multiply the detector area with the Be transmission and store this as the ancillary response function (ARF).

Obtaining the background is more involved. For very strong sources, such as the Crab, there is some transmission seen through the aperture stop itself and this manifests as a much fainter secondary `ring' outside the stray-light (seen as a light blue color in Figure \ref{detplots}). Backgrounds can therefore not be taken from the region adjacent to the stray-light. Fortunately not all observations had stray-light on them; the orientation of the spacecraft ensured that the optical bench was blocking the module for some observations, and these are marked in Table \ref{obsid} with `bkg'. Because of the solar activity during the 2015 observations, which was absent in 2016, we cannot use backgrounds from 2015 for spectra from 2016. We obtain clean backgrounds for all detectors on FPMA for both epochs, but we only have a clean background for FPMB from epoch 2016. For FPMB epoch 2015 we had to make use of \texttt{nyskybkg} \citep{Wik2014} on obsID 10110002002. We show the detector plot in Figure \ref{detplots} (bottom right); there is transmission through a section of the optical bench contaminating most of the module, though not all. We follow the approach outlined in the \texttt{nuskybkg} guide of extracting as much clean background as possible from the regions outlined by the white polygons. We run these two spectra through \texttt{nuskybkg} and fit the background, thereby inferring what the true background is for the rest of the detector.

The focused observation, 10002001009, was reduced using \texttt{nupipeline} CALDB version 20160606. We extracted using a 200\as\ radius circular region, taking a background as close as possible without including any source photons, and generated spectra and response files with \texttt{nuproducts}. We used default parameters throughout.

\begin{table}
\centering
\caption{Crab results}
\begin{tabular}{l||c|c|c}
\hline
Year & $\Gamma$ & Normalization & Flux$_{3-50\,\mathrm{keV}}$ \\
 & & & ($10^{-8}$ erg cm$^{-2}$ s$^{-1}$) \\
\hline
2015 & 2.098$\pm$0.006 & 9.52$\pm$0.19 & 3.379$\pm$0.014 \\
2016 & 2.116$\pm$0.007 & 9.91$\pm$0.20 & 3.353$\pm$0.014 \\
Both & 2.106$\pm$0.006 & 9.71$\pm$0.16 & 3.368$\pm$0.011 \\
\hline
\hline
Module &	 Detector & Pt & CZT \\
& & ($\mu$m) & ($\mu$m) \\
\hline
A & 0 & 0.094$\pm$0.006 & 0.218$\pm$0.039\\
A & 1 & 0.089$\pm$0.007 & 0.304$\pm$0.041\\
A & 2 & 0.074$\pm$0.006 & 0.370$\pm$0.040\\
A & 3 & 0.094$\pm$0.008 & 0.329$\pm$0.043\\
\hline
B & 0 & 0.108$\pm$0.004 & 0.187$\pm$0.026\\
B & 1 & 0.066$\pm$0.005 & 0.292$\pm$0.031\\
B & 2 & 0.079$\pm$0.012 & 0.270$\pm$0.071\\
B & 3 & 0.067$\pm$0.006 & 0.282$\pm$0.036\\
\hline
\end{tabular}
\label{result}
\end{table}

\begin{figure}
\begin{center}
\includegraphics[width=0.4\textwidth]{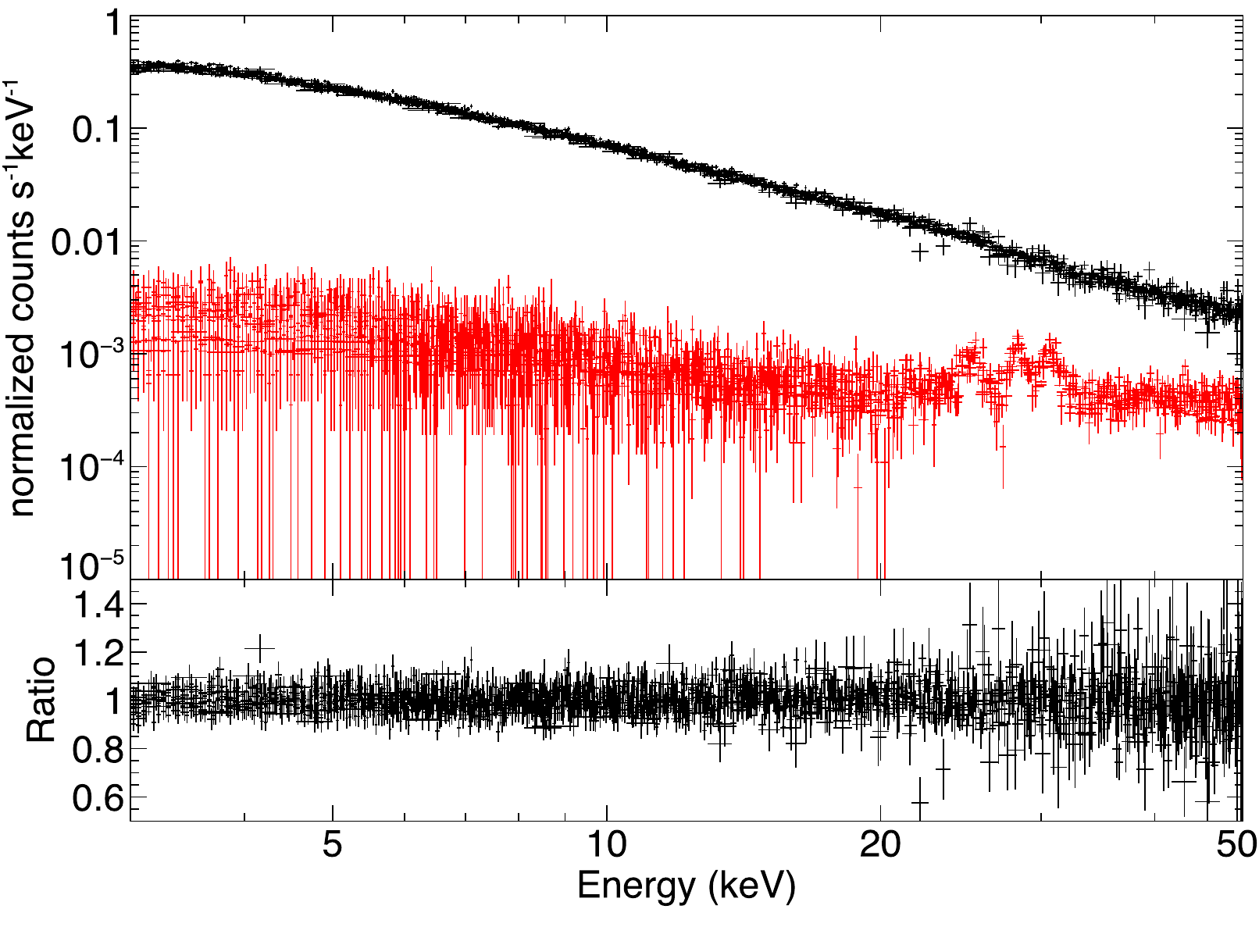}
\end{center}
\caption{Best fit of all 14 spectra from both epochs with backgrounds (red).}
\label{powfit}
\end{figure}

\begin{figure}
\begin{center}
\includegraphics[width=0.4\textwidth]{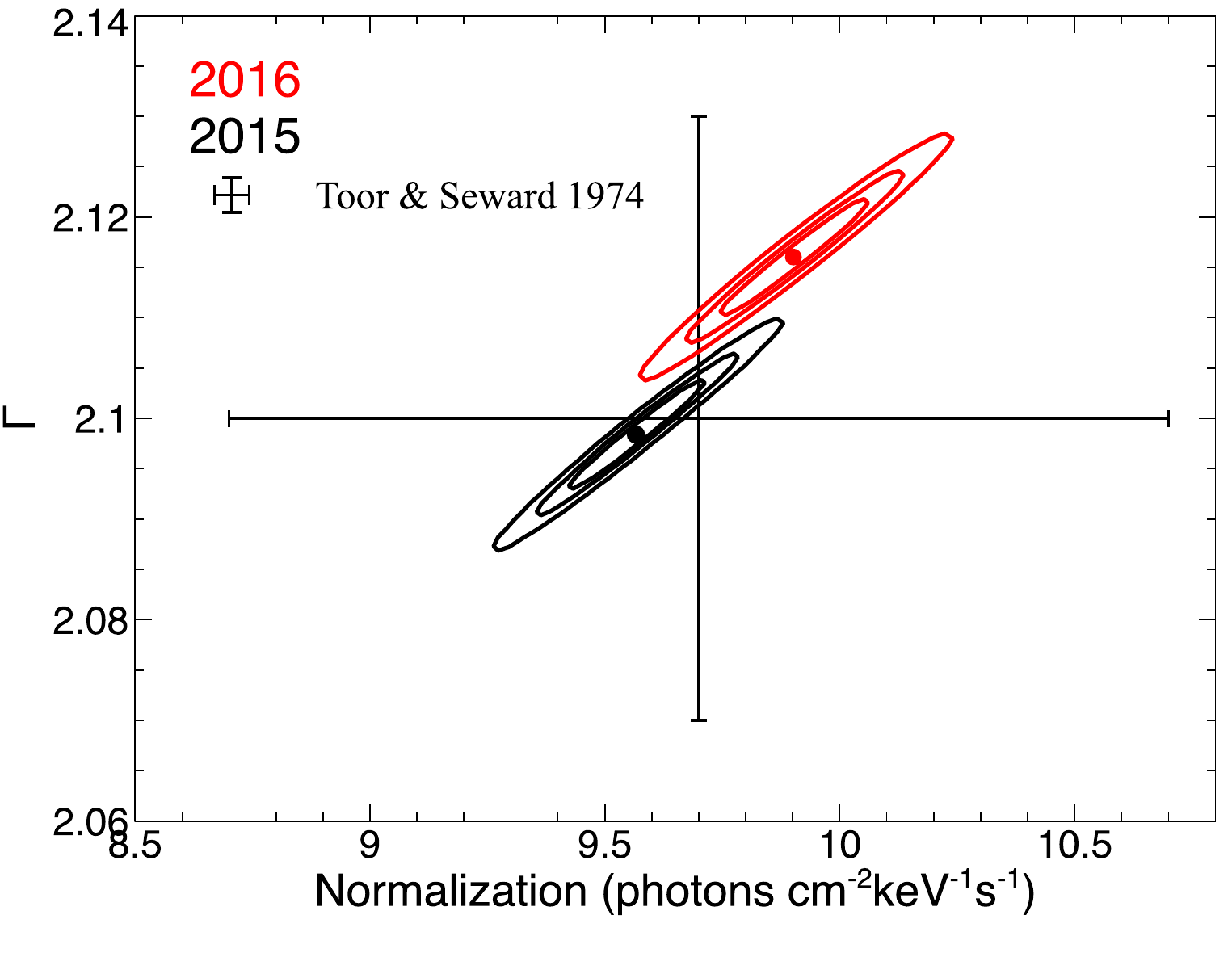}
\end{center}
\caption{Contour plots of the normalization and slope of the two epochs.}
\label{powcontour}
\end{figure}

\section{Results}

We use an XSPEC model \texttt{nuabs$\times$tbabs$\times$pow} to represent the Crab spectrum. The model \texttt{nuabs} is an absorption model for the detector with cross-sections created by GEANT4 \citep{Agostinelli2003}. The adopted photon interaction is the Livermore low-energy EM model based on the evaluated photon data library, EPDL97 \citep{Cirrone2010}. The model has four parameters: the thickness of Pt, CZT (CdZnTe), and Zn, and the Cd ratio. We only fit for Pt and CZT and keep the other two frozen to Zn=0 and Cd ratio=0.9. For \nh\ we use \citet{Wilms2000} abundances and \citet{Verner1996} cross-sections. Attempts at letting \nh\ remain unbound resulted in the value being larger than 2$\times 10^{22}\mathrm{cm}^{-2}$, which is ten times what is expected and can be ruled out. The reason for the high value is a degeneracy with the detector absorption parameters, which consequently took unlikely values. The \nh\ of the Crab is sensitive to individual instrument calibrations, but measurements from several observatories constrain it to lie in the range from 2 to 6$\times 10^{21}\mathrm{cm}^{-2}$, with an average value of $\sim 4 \times 10^{21}\mathrm{cm}^{-2}$ \citep{Kirsch2005}. In the original calibration we measured the Crab column to be \nh=$2.2 \pm 2.0 \times 10^{21}\mathrm{cm}^{-2}$, and since we do not want to introduce too many changes and it is within the measured range, we maintain this value. At 4\,keV the absorption of this column is 1\% and if the column was increased to $4 \times 10^{21}\mathrm{cm}^{-2}$ the absorption at 4\,keV would be 2\%. With the best fit detector absorption parameters frozen, \nh\ has for these observations a 90\% confidence limit of $\pm 1.1 \times 10^{21}\mathrm{cm}^{-2}$.

We fit all 14 data sets (eight for 2015 and six for 2016) simultaneously in \texttt{XSPEC} \citep{Arnaud1996} using C-stat fitting statistics \citep{Cash1979} and show the fit and ratio residuals in Figure \ref{powfit}. The spectra can not be combined due to the differences in the RMFs for each detector, and because the Crab spectrum could potentially be different between the two epochs. We thus allow the slope and normalization to differ between 2015 and 2016, but require that the detector absorption parameters remain the same for each detector in both epochs. Since we were unable to take backgrounds directly from the same observations, and had to model them for FPMB epoch 2015, we limit the fit to be between 3--50\,keV to reduce the influence of a possible bad background. At 50\,keV the background is an order of magnitude below the source.

The values of the detector absorption parameters are listed in Table \ref{result}. As compared to the values reported in \citet{Madsen2015} the CZT dead-layer thickness is $\sim 50\%$ higher while the Pt thickness is $\sim 50\%$ lower. We do not believe this difference to be due to a contamination effect, but instead is the result of untangling the mirror response from the true absorption. Changing the Crab column to \nh$=6\times 10^{21}\mathrm{cm}^{-2}$, which we consider an upper limit, changes the Pt and CZT values by $\sim 5\%$ and is comparable to the error of the thicknesses themselves. 

We summarize the Crab spectral results in Table \ref{result} and in Figure \ref{powcontour} show the 1, 2, and 3$\sigma$ contours of the Crab normalization and slope for the two epochs. The intrinsic flux measured between 3--50\,keV for the two epochs is practically identical, $F_{2015} = 3.379\pm 0.014 \times 10^{-8}$\,erg\,cm$^{-2}$\,s$^{-1}$ and $F_{2016}=3.353\pm 0.014 \times 10^{-8}$\,erg\,cm$^{-2}$\,s$^{-1}$, with the difference on the order of the errors. In contrast, the slopes between the two epochs have a significant offset, $\Gamma_{2015}=2.098\pm0.006$ and $\Gamma_{2016}=2.116\pm 0.007$, which results in a difference in normalization of $N_{2015}=9.52\pm0.19$ and $N_{2016}=9.91\pm0.20$. In both slope and normalization there is overlap at the 2$\sigma$ level, and so the measurements are consistent at 3$\sigma$. If we were to assume the Crab spectrum to be the same for both epochs, the best fit finds $\Gamma=2.106\pm 0.006$, $N=9.71\pm 0.16$ and $F(3-50\mathrm{\,keV})=3.368\pm 0.011 \times 10^{-8}$\,erg\,cm$^{-2}$\,s$^{-1}$. 

The simplicity of the instrumental response allows us to place a tight limit on the flux. The ARF is just the Be absorption, which is known to 1\% from lab experiments, and the detector area precisely calculated as the area of the polygons. The detector response, RMF, has been generated using a charge transport model customized to the \nustar\ hybrid design \citep{Kitaguchi2011}, and for this type of flat spectrum the errors in the line spectrum do not matter. The quantum efficiency (QE) of the detectors is 98\% between 4--40\,keV and understood to $<$1\%. If we allow errors of 1\% on both the RMF and ARF respectively, and another 1\% for calculating the detector area and uncertainties in column, we have a 3\% systematic error on the intrinsic flux in addition to the 90\% confidence on the intrinsic flux of 0.4\%. Since changes in flux can come from slope changes, normalization changes, or changes in both, it is not possible to say how the systematic errors affect the individual parameters of $\Gamma$ and $N$ without knowing exactly how the errors in the responses look as a function of energy. However, if we were to assume that there are no slope changes but only a normalization change, then the systematic error would directly apply to the normalization, which has been measured at 90\% confidence to 2\%.

Comparing to \citet{Toor1974}, the \nustar\ epoch averaged values of $\Gamma=2.106\pm 0.006$ and $N=9.71\pm 0.16$ are in excellent agreement with their $\Gamma=2.10\pm 0.03$ and $N=9.7\pm 1.0$. This supports the findings of \cite{Wilson2011} and \cite{Shaposhnikov2012} that the flux changes observed on yearly timescales from the Crab vary about a steady mean rather than a long term decreasing (or increasing) trend. As of yet, there is no clear understanding what might cause these yearly variations of a few percent, but it has been proposed that they are tied to the Gamma-ray flares observed in the Crab by \textit{Agile} \citep{Tavani2011} and \textit{Fermi} \citep{Abdo2011}. In this scenario the flux variations are due to the afterglow of the flares as the high-energy electrons are advected through the synchrotron nebula and cool via synchotron losses \citep{Cerutti2013,Kroon2016}. 


In the above we have assumed that the phase-averaged integrated Crab spectrum of the nebula and pulsar can be approximated by a powerlaw. However, spatially resolved spectroscopy of the Crab with \chandra\ and \nustar\ has shown that the spectral shape is changing across the remnant. Using \chandra\ data \citet{Mori2004} find that the spectra below 10\,keV can be fitted with powerlaws of varying index, while using \nustar\ data \citet{Madsen2015b} measured clear breaks of these powerlaws at $\sim$10\,keV with increases in slope of $\Delta\Gamma\sim 0.1-0.2$ confined to the torus feature of the nebula. Additionally, the broadband pulsed spectrum has been found to be curving \citep{Kuiper2001, Madsen2015b}. The Crab is dominated by its nebular spectrum, but even then the superposition of all these disparate spectra should not mathematically add up to another powerlaw. \citet{Weisskopf2010} investigated if a deviation from a curved spectrum could be measured with current instrumentation for two different models for the integrated nebula and pulsar Crab spectrum. They concluded that for one model it would be possible and that the \rxte\ spectrum already excluded this model. For the other model they concluded that even if the instrument responses were perfectly known it would be difficult. \nustar\ was not included in this investigation, and with its broader energy band it may be possible to measure a deviation from a powerlaw. We did not attempt to fit the models from \citet{Weisskopf2010}, but we investigated if a better fit could be achieved with a curved spectrum like a (in XSPEC notation) \texttt{logpar}, \texttt{brokenpowerlaw}, or \texttt{cutoffpowerlaw}. We also measured the spectral slope in smaller bands, but in neither case did we find improvement, or a significant deviation in spectral slope from the measured broadband powerlaw slope. We thus conclude that in the collimated configuration over the 3--50\,keV band, \nustar\ is not able to measure a deviation from a powerlaw, if it exists, in the current observations.

Finally, we compare to the focused Crab observation taken in 2016 together with the stray-light campaign. \nustar\ was calibrated against a Crab of $\Gamma=2.1$ and $N=8.7$ with the choice of normalization set in order to have agreement with the contemporary observatories, \chandra, \swift, \suzaku, and \xmm\ \citep{Madsen2015}. The best fit focused observation gives $\Gamma=2.098 \pm 0.001$, $N=8.44 \pm 0.02$ and $F(3-50\mathrm{\,keV})=2.990 \pm 0.003 \times 10^{-8}$\,erg\,cm$^{-2}$\,s$^{-1}$. Formally the fitting errors are very good. However, due to the uncertainty in the optical axis location, the absolute errors are $\Delta \Gamma =0.01$ and $5\%$ on the flux. This brings the slope in agreement between the focused and stray-light Crab observations, but leaves the flux $\sim 12\%$ lower, which is what is expected from the calibration.

Inspecting the current values of cross-normalizations relative to \nustar\ in the limited energy band 3--7\,keV from \citet{Madsen2015} we have: $C_{Chandra/\mathrm{HEG}}$=1.10, $C_{Swift/\mathrm{XRT}}$=1.05, $C_{Suzaku/XIS}$=0.95, and $C_{XMM-Newton/\mathrm{MOS}}$=1.0. The observatory currently closest to the true intrinsic absolute flux of the Crab in the 3-7\,keV band is therefore \chandra. We stress, though, that this does not inform about the slopes of the respective instruments, just the integrated flux in the limited band.


\section{Conclusion}
We have presented the analysis of stray-light observations of the Crab. In this configuration \nustar\ acts as a collimated instrument and is particularly simple in terms of the instrument response. We have measured the intrinsic absolute flux of the Crab to better than $4\%$, where we have conservatively added a systematic error of 3\%. We measure the spectral parameters of the Crab in two different epochs and find that while the flux remains steady to within 1\%, the slope and normalization are slightly different. Both values are in excellent agreement with the measurements done by \citet{Toor1974} 42\,years prior and indicate that the observed variability of the Crab is not part of a long term trend, but fluctuations around a steady mean. 

The true intrinsic flux of the Crab, as measured by the stray-light, is 12\% higher than that measured through the \nustar\ optics. This is understood because of the spectral parameters ($\Gamma=2.1$, $N=8.7$) used to calibrate the mirror response in \citet{Madsen2015}. The slope is in agreement.

We were able to measure new detector absorption parameters and separated out the mirror response from the previous observation. We have as a result updated the detector absorption files for CALDB version 20160606. At the present time, there is no plan to adjust for the 12\% absolute flux difference in the mirror response.

\acknowledgments
We thank the Referee for the helpful comments and suggestions. This work was supported under NASA Contract No.
NNG08FD60C, and made use of data from the \nustar\ mission,
a project led by the California Institute of Technology,
managed by the Jet Propulsion Laboratory, and funded by the
National Aeronautics and Space Administration. We thank
the \nustar\ Operations, Software and Calibration teams for
support with the execution and analysis of these observations.
This research has made use of the \nustar\ Data Analysis
Software (NuSTARDAS) jointly developed by the ASI Science
Data Center (ASDC, Italy) and the California Institute
of Technology (USA).

{\it Facility:} \facility{CXO}, \facility{NuSTAR}, \facility{Swift}, \facility{Suzaku}, and \facility{XMM}

\bibliography{bib}
\bibliographystyle{jwapjbib}

\end{document}